\documentclass[12pt]{article}
\usepackage{graphicx}

\usepackage{setspace}
[1999/03/29 PSNFSS v.7.2
Times font as default roman
: S Rahtz]

\begin{document}

\renewcommand{\thefootnote}{\fnsymbol{footnote}}
\sloppy
\newcommand{\rp}{\right)}
\newcommand{\lp}{\left(}
\newcommand \be  {\begin{equation}}
\newcommand \ba {\begin{eqnarray}}
\newcommand \bas {\begin{eqnarray*}}
\newcommand \ee  {\end{equation}}
\newcommand \ea {\end{eqnarray}}
\newcommand \eas {\end{eqnarray*}}

%\title{Linking the Burridge-Knopoff model of earth quakes and the capital asset price model of financial markets}
%\title{Measuring ``price-quakes'' in the network of the world's stock exchanges}
\title{\large Contagion in the world's stock exchanges seen as a set of coupled oscillators}
\thispagestyle{empty}
\newpage

\maketitle
\author{. \\
Lucia Bellenzier$^1$, J\o rgen Vitting Andersen$^2$,  
and  Giulia Rotundo$^3$ \\
} \vskip 24pt
 $^1$Department of Statistics and Quantitative Methods,
		University of Milano-Bicocca,
Piazza dell'Ateneo Nuovo 1, 20126 Milano, Italy. 
email: {lucia.bellenzier@unimib.it} \\
 $^2$
CNRS, Centre d\'{}Economie de la Sorbonne,  
Universit\'e Paris 1 Panth\'eon-Sorbonne,  
Maison des Sciences Economiques,
106-112 Boulevard de l'H\^opital 75647 Paris Cedex 13, France.
email: Jorgen-Vitting.Andersen@univ-paris1.fr, tlp.: +33 (0)144078263 fax: +33 (0)144077676
\\
%\email{Jorgen-Vitting.Andersen@univ-paris1.fr}
 $^3$  
Department of Methods and Models for Economics, Environment and Finance,
Sapienza University of Rome, 
via del Castro Laurenziano 9, 
00161 Rome, Italy. email: giulia.rotundo@uniroma1.it

\begin{abstract}
%Synchronization which happens through so-called integrate-and-fire (IAF) oscillators are thought to be relevant in a large %class of phenomena in Nature, ranging from how swarms of fireflies coordinate their flashes, to the coordination in neural %activity which can lead to epilepsy. 
{\bf Abstract} \\
We study how the phenomenon of contagion can take place in the network of the world's stock exchanges due to the behavioral trait "blindeness to small changes". On large scale individual, 
the delay in the collective response may significantly change the dynamics of the overall system.
We explicitely insert a term describing the behavioral phenomenon in a system of equations that describe the build and release of stress across the worldwide stock markets. In the mathematical formulation of the model,
each stock exchange acts as an integrate-and-fire oscillator. Calibration on market data validate the model. 
 One advantage of the integrate-and-fire dynamics is that it enables for a direct identification of cause and effect of price movements, without the need for statistical tests such as for example Granger causality tests often used in the identification of causes of contagion. Our methodology can thereby identify the most relevant nodes with respect to onset of contagion in the network of stock exchanges, as well as identify potential periods of high vulnerability of the network. The model is characterized by a separation of time scales created by a  slow  build up of stresses, for example due to (say monthly/yearly) macroeconomic factors, and then a fast (say hourly/daily) release of stresses through "price-quakes" of price movements across the worlds network of  stock exchanges.  
\end{abstract}

\vspace{1cm}
\noindent


\section{\large 1. Introduction}
The financial market turmoil around the 2008 sub-prime crisis has been an awakening for academics and policy-makers to understand and capture the linkages and vulnerabilities of the financial system. Much of such efforts have been focused around systemic risks and contagion phenomena. The issue of instability is however not new and has been put forward since the great depression era of the 1930's by e.g. Fisher (1933) and Keynes (1936). The subject itself is nonetheless  not without controversy since some argue that the use of the term contagion is misplaced and financial markets rather show a high level of market co-movement at all times and should rather be called  market ``interdependence'' (Forbes and Rigobon 2002).

A large part of studies on contagion relates to correlation-based networks. For instance, in Chiang et al. (2007), the contagion is detected from the statistical analysis of the correlation among markets and with a behavioral perspective that consists in
the interpretation of the continued high correlation as herding. As mentioned in Aloui et al. (2011) studies of the transmission of return and volatility shocks
from one market to another as well as studies of the cross-market
correlations are essential in finance, because they have many
implications for portfolio allocation. Their paper used a multivariate copula approach to examine
the extreme co-movement across markets in order to study the harmful consequences of contagion
effects on portfolio selection. 

In Bekaert et al. (2003), the contagion is defined as correlation between markets in excess of what would be implied by the
fundamentals. However, this definition makes the measurement quite difficult because there is no common agreement
on the definition of fundamentals, although the model may explicitly consider macroeconomic variables
(Syllignakis and Kouretas (2011)).
In Caporale et al.(2005), Chiang et al. (2007), the contagion is detected through co-movements of the correlation. Thus, the
main issue is on modeling the correlations Celik (2012), Dimitriou et al. (2013), Gjika and Horvath (2013), Mensi et al. (2013) or the cointegration
Hong et al. (2009). Such debates led to the discussion of the difference between interdependence and contagion, Ahmad et al. (2013),
Aloui et al. (2011), Corsetti et al.(2005).
In Bae et al. (2003) it was proposed to consider contagion as  a
phenomenon associated with extreme returns: if there is contagion, small return
shocks propagate differently from large-return shocks. In their study they focused on counts of coincidences
of extreme returns rather than on correlations of joint extreme
returns. The different role of propagation of small versus large returns will be seen to be a key ingredient in our model and indeed to be one of the main mechanisms behind the creation of contagion.

Another issue relates to cause and effects in contagion. For example a study of Yang J., \& Bessler D. A. (2008) was able to use a vector auto-regression analysis to pinpoint that the 1987 crash originated in the US markets whereas a following upward movement of the Japanese market was important for the subsequent recovery. However a clear-cut conclusion of what started the market turmoil and what made it end is often difficult. For example Roll (1988) came to a different conclusion in his analysis  of 23 of the major markets worldwide, and argued that in fact the international stock market crash of 1987 started in
Asian countries, other than Japan, and from there spread to Europe, the US and finally reached Japan. A different way to obtain information about cause and effect is through surveys. In Shiller (1989) a survey places the US as playing the dominant role in the international 1987 crash.

Our study offers to take a new look with respect to network analysis of contagion by introducing a model in which cause and effect is inherently defined without need for statistical test such as for example Granger causality tests. We will address new issues with respect to  network analysis in order to get a statistical understanding of the pathology of contagion and  also consider the question of cause of effect, something which the structure of the model allows a direct identification of without the need for statistical tests on correlations.

\begin{figure}[h]
%\includegraphics[width=14cm\textwidth, natwidth=610, natheight=642]{figure_world_map.ps}
%xxx 
\includegraphics[width=10cm]{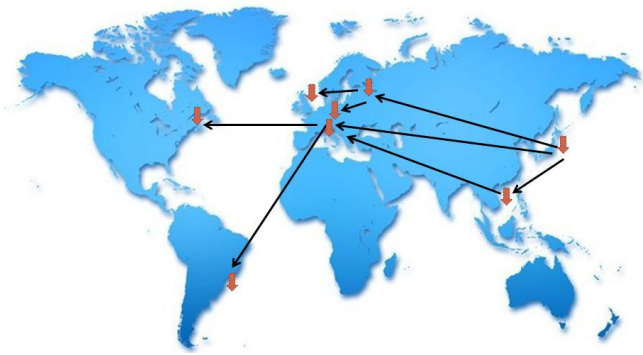}
\caption{\protect\label{Fig1}
``Price-quake''. One of the main advantages of the non-linearities in the integrate-and-fire oscillator model is that it enables for a clear-cut identification of cause and effect. The figure illustrates one example of a price-quake following an initial minus seven ercentage price move of the Japanese stock market on the 23/05/2013. 
}
\end{figure}

In the following we will suggest to consider the world's network of stock exchanges as a network of coupled oscillators. The idea is to consider each exchange as an oscillator of a ``force field'' (to be defined below) which can influence the other oscillators in the network. Our methodology enables a new understanding of how impact generated through non-linear price dynamics can propagate across markets and can be used to study the origins behind contagion effects in the network of stock exchanges. Contagion can in such picture be seen as synchronization of the network of stock exchanges as a whole, caused in a large part of the exchanges which adjust their ``rhythms'' (by pricing in price movements of the other exchanges) thereby producing  a global aggregate signal. One of the main features of our model is a separation of time scales with a slow price dynamics due to economic fundamentals for a given country, and a fast price dynamics due to impact across markets. An example illustrating how the method identifies the network of propagation after a large stock movement of the Japanese stock market the 23/05/2013 is shown in figure~\ref{Fig1}

\section{\large 2. Empirical methodology}

\subsection{\large 2.1 Models of coupled Integrate-And-Fire (IAF) oscillators}
\begin{figure}[h]
 \includegraphics[width=14cm]{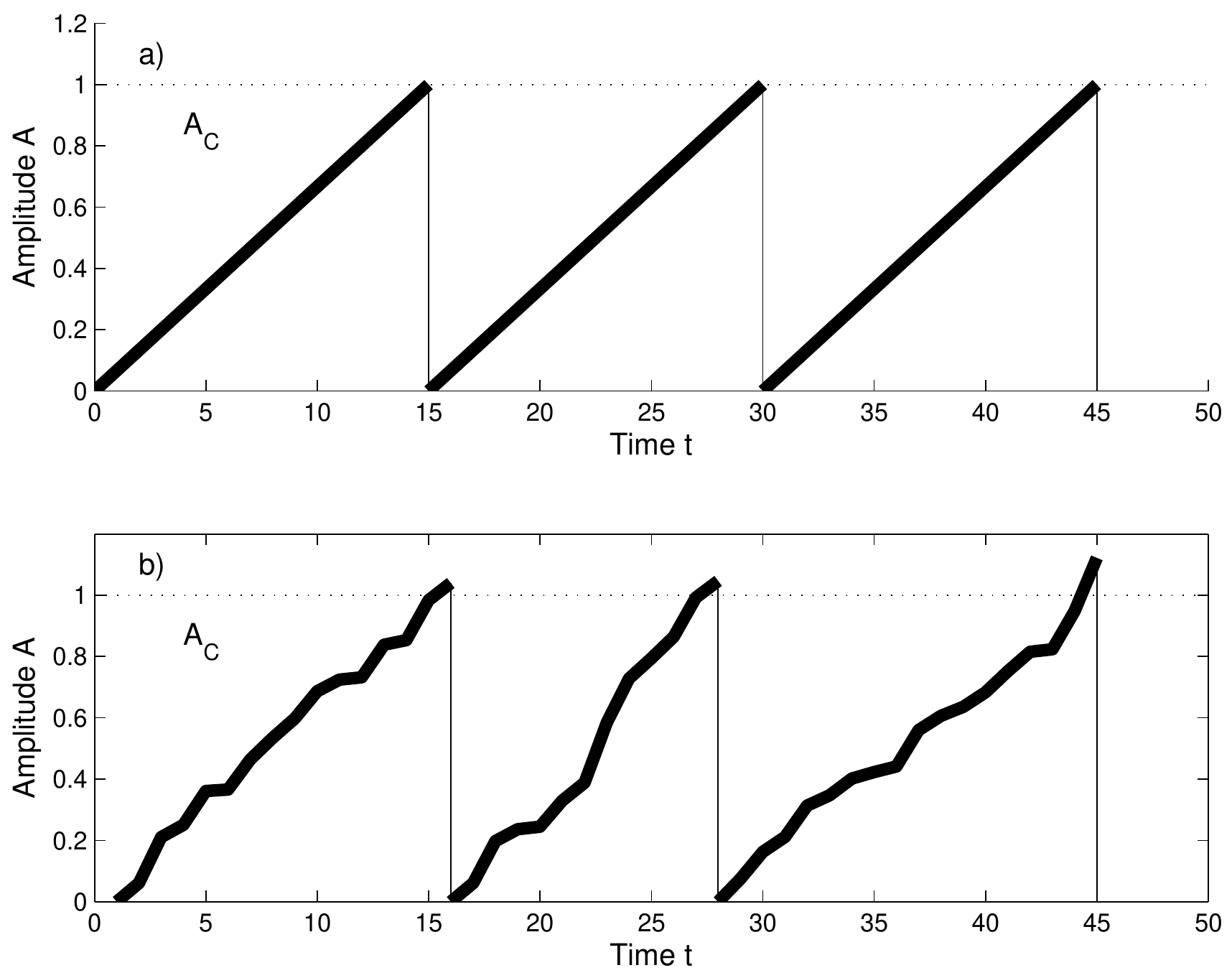}
\caption{\protect\label{Fig2}
Illustration of an IAF oscillator. a) illustrates the case where the amplitude $A(t)$ of an IAF oscillator integrate linearly in time until it reaches a critical value $A_C$ after which it discharges by setting $A(t) = 0$. The case in a) can be seen as one IAF oscillating over three periods of time, or equivalent three identical and uncoupled IAF oscillators oscillating over one period of time. b) shows over three time periods an IAF oscillator having a random frequency, or equivalently, three different unit oscillators with random frequency over one time period.
}
\end{figure}

In order to illustrate the concept of an IAF oscillator consider figure~\ref{Fig2}a. The figure shows an oscillator with an amplitude, $A(t)$, which increases constantly versus time $t$ until it reaches a threshold $A_C=1$ after which it discharges and its amplitude $A$ is reset to zero. The process then repeats until a new discharge takes place and so forth and so on. Figure~\ref{Fig2}a could equally well be seen as three identically and independent IAF oscillators  operating with the same constant frequency. Assuming furthermore independence between the different units of oscillators the system of oscillators is trivially described by the same oscillation as a single unit oscillator. Figure~\ref{Fig2}b illustrates over three time periods an IAF oscillator having a random frequency, or equivalently, three different unit oscillators with random frequency over one time period. In this case the aggregate response of a system with several units is less trivially especially if there exists a coupling, i.e. dependence, between the units.

As will be seen in the following sections we argue that each stock exchange can indeed be seen as an unit IAF oscillator, with oscillations determined by two contributions: i) a contribution from a characteristic proper (random) frequency, and ii) another contribution determined by the strength of coupling to the other unit oscillators in the network of stock exchanges.

\subsubsection{\large 2.1.1 The Olami-Feder-Christensen (OFC) model}
In order to introduce the reader to the dynamics of coupled IAF oscillators we first discuss the Olami-Feder-Christensen model (Olami, Feder and Christensen, 1992) since it can be seen as a special case of our more general IAF oscillator model for the network of the world's stock exchanges. The introduction of the OFC model will allow a simple but general description of the special non-linear dynamics of activity which is the hallmark of IAF oscillator models and which will also be seen to lead to a non-linear price dynamics of the world's stock exchanges. 
Originally the OFC model was introduced as a model of earth quake activity to capture the stick-and-slip dynamics seen in earth tectonic plate movement. As will be seen such stick-and-slip motion appears in our model due to ``news'' in terms of the price movements of other stock exchanges which first imposes stresses across markets which then subsequently become priced in. "Stick" will be seen corresponding to the build up of stresses imposed on one stock exchanges due to the price movement of another stock exchange, and "slip" corresponds to the following release of such stresses when they have been taking into account, i.e. have been priced in.

The OFC model is described in terms of a force variable, $F_{i,j}$ defined on a discrete two-dimensional set of blocks (the set of blocks representing an earth tectonic plate) and given by the coordinates $(i,j)$. The dynamics of the model can be represented by a cellular automaton description via the two rules I), II) presented in the following. The rules determine the non-linear dynamics of the OFC model and will re-appear, but in a more complex form,  when we in the following introduce the price dynamics of the world's stock exchanges: \\

%\begin{center}
%\begin{numerize}

%\item 
{\em
I) A critical site $(i,j)$ is defined as a site which has the force larger than a certain magnitude $F_C$, i.e. $|F_{i,j}| > F_C$. If there are no critical sites (as e.g. when the model is first initialized) then first find site $(i^*,j^*)$ with maximal stress: 
\be
\label{F_max}
F^{max}_{i^*,j^*} =  max_{(i,j)} F_{i,j}
\ee
Add the same additional stress $\eta_{i^*,j^*} \equiv (F_C - F^{max}_{i^*,j^*})$ on all sites so that the site with maximal stress, $(i^*,j^*)$, now becomes 
critical:
\be
\label{F_max_add}
F_{i,j}  =  F_{i,j} +  \eta_{i^*,j^*} \ \  \forall (i,j) 
\ee
}

{\em

%\item 
II) Else if there are $N$ critical sites $(i_1^*,j_1^*), (i_2^*,j_2^*), ..., (i_N^*,j_N^*)$ 
then {\bf simultaneously} release the stress of all those sites (make them ``topple''):
 
\be 
\label{F_relax}
F_{i^*,j^*}  =  0 \ \ \forall (i^*,j^*) 
\ee

and transfer a certain fraction, $\alpha$, of their stress, $F_{(i^*,j^*)}$ (prior to the toppling) to their nearest neighboring  sites $(i^*_{NN},j^*_{NN})$:
\be 
\label{F_topple}
F_{i^*_{NN},j^*_{NN}} =  F_{i^*_{NN},j^*_{NN}} + \alpha F_{(i^*,j^*)} \ \ \forall (i^*_{NN},j^*_{NN}) 
\ee
}
\begin{figure}[h]
 \includegraphics[width=14cm]{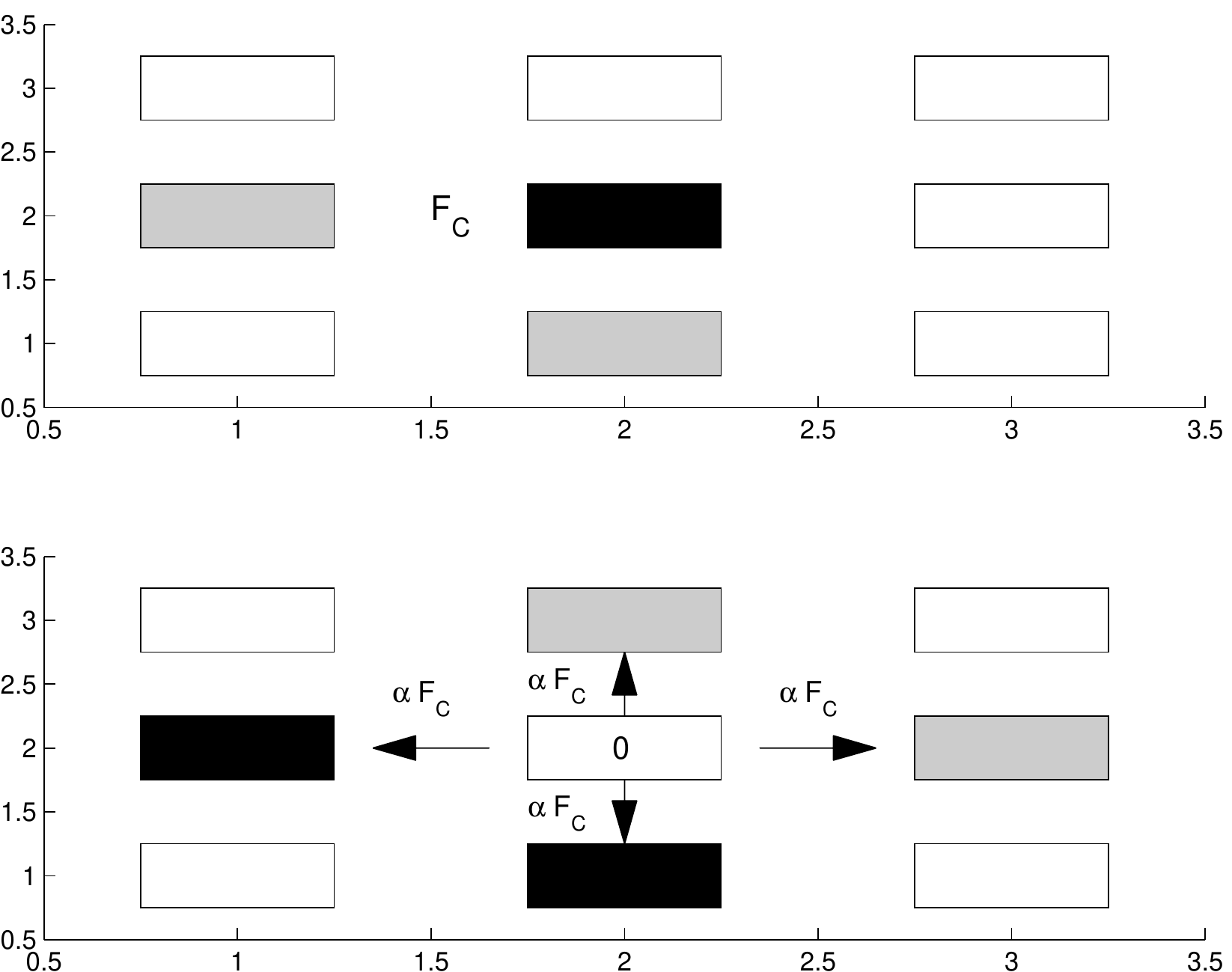}
\caption{\protect\label{Fig3}
Illustration of the force dynamics in the spring-block model of OFC model. a) The central block has exceed the critical value, $F_C$, indicated by the black color of the block. Neighboring blocks are all sub-critical (indicated by the gray or white color). b) the stress acting on the central block is released (the force reset to 0) and redistributed to neighboring blocks which subsequently become either critical or sub-critical. The avalanche of force re-distributions then continue by releasing the stresses on the two new block which have passed the critical value which might lead to the creation of yet other critical blocks and so on and so forth. 
}
\end{figure}
One way to illustrate the non-linear dynamics behind Eqs.(1)-(4) is the representation of the update of a critical block/site shown in Figure~3a-b. Figure~3a illustrates the event corresponding to rule I where there were no critical blocks/sites in the network and where the block in the middle had the largest stress (among all the blocks in the network) acting on it. According to Eq.(2) a constant amount of stress was added to all the blocks such that the center block in figure~3a attained the critical threshold $F_C$ which is indicated by the black color of the center block. 
%On can illustrate graphically the dynamics behind the two rules as seen in figure~\ref{Fig3}.  
%Figure~3a shows the case with a block, shown in the middle, which has a maximal stress, see Eq.(1). The stress on this block was increased until it became critical, i.e. it attained the value $F_C$, which is indicated by the black color. The given increase in stress was added to all the blocks in the system according to Eq.(2). 
Figure~3b illustrates how the stress of the critical block is redistributed to its neighbors, leading to two new critical blocks, according to Eq.(4). The stress of the critical block was then reset to 0, as expressed from Eq.(3).

From Eq.(1)-(4) and Figure~3a-b one can now see that the OFC model corresponds to a network of IAF oscillators. Take for example the central block shown in figure~3a: this block has experienced subsequent increases of the amplitude of stress acting upon it, similar to the IAF oscillator in Figure~2b.  Once the force on the block hits the threshold $F_C$, it is reinitialized via Eq.(3) in analogy to the dynamics of the IAF oscillator in Figure~2b. The coupling between the different IAF oscillator units happens through Eq.(4). Such a coupling was shown (Olami, Feder and Christensen, 1992) to lead to a dynamics with power law probability distribution functions in the size of certain events called ``avalanches'', with ``avalanche'' defined by  the number of sites involved in a disturbance as described via Eq.(2) under rule II. Having seen how the OFC model can be understood as a coupled IAF oscillator network, we next turn our attention to our ``price-quake'' model.

%\end{enumerate}
%\end{center}

\subsubsection{\large 2.1.2 The ``price-quake'' oscillator model}
To see how the idea of IAF oscillators can be defined in a financial market setting we consider in the following the  worlds stock exchanges as a network where each exchange can influence the price dynamics of the other exchanges. Small price changes of a given exchange are not assumed to have an immediate impact for other markets, only larger, eventually aggregate price moves over several time periods, will be taken to have an impact across markets. Here we have used the tendency for humans to ignore minor events and only react to larger stimuli, a fact  known in Psychology as ``change blindness'' (Jones, Crowell, \& Kapuniai 1969; Lewin, Momen, Drifdahl, \& Simons 2000; Rensink 2002). 
In a financial market context such a phenomenon has  been documented through empirical analyses
De Bondt and Thaler (1985), Lin (2012), Lin S., and Rassenti S. (2012). Such behavioral responses allows to describe financial models through different perspectives  H\"am\"al\"ainen et al. (2013); Vitting Andersen and Nowak (2013). It will also be assumed that the relative capitalization of two markets as well as their geographically closeness affect the magnitude of the impact one market can have on another. 

The ``price-quake'' model of Vitting Andersen et al. (2011) can be formalized as follows: at time $t$, a trader of a given stock  exchange $i$ estimates the  price $P_i(t)$ of the index as $P_i(t)=P_i(t-1)\exp{(R_i(t))}$ with $R_i (t)$ the return of stock exchange $i$ between time $t-1$ and time $t$. Taking the trading volume as proxy for the relevance of reaction to new information, and noting that the trading volume is highest around the open/close of a market, we subsequently only take into account the opening or closing prices of the different indices. If $t$ denotes say the close of index $i$, $R_i(t)$ will therefore denote  the return of index $i$ between the open of index $i$ at time $t-1$ and close at time $t$. The return $R_i(t)$ is calculated by the traders taking into account price movements in other stock exchanges (that happened in the time between the opening-close or close-opening of index $i$) as well as local economic news, $\eta_i (t)$ relevant only for the stock exchange $i$: 
\be
\label{R_return}
R_i (t) = {1 \over N_i^*} \sum_{j\neq i}^N \alpha_{i,j} 
\theta (|R_{ij}^{cum} (t-1)| > R_C) \times R_{ij}^{cum} (t-1) \beta_{i,j} + \eta_i (t)
\ee
\be
\label{R_cum1}
R_{ij}^{cum} (t) = [1 - \theta (|R_{ij}^{cum} (t-1)| > R_C) ] \times R_{ij}^{cum} (t-1) + R_j(t)\ \ \forall j \neq i
\ee
%\be
%\label{R_cum2}
%R_{ji}^{cum} (t) = R_{ji}^{cum} (t-1) +  R_{i} (t) \ \ \forall j \neq i
%\ee
\be
\label{N_*}
N_i^* =  \sum_{j\neq i}^N  
\theta (|R_{ij}^{cum} (t-1)| > R_C) , \ \ 
\alpha_{i,j} = 1 - \exp{(-K_j/(K_i \gamma))},  
 \ \ 
\beta_{i,j} = \exp{(-(|z_i-z_j|)/\tau)}
\ee
$R^{cum}_{ij}$ describes the stress that exchange $j$ imposes on exchange $i$, see second term in Eq.(6) and first term in Eq.(5). 
Compared to the pricing of individual stocks, the structure of Eq.(\ref{R_return}) has similarities to the pricing obtained via the Capital Asset Price Model (CAPM) (Treynor, 1999; Lintner 1965; Sharpe 1964) since it determines how a given stock exchange, $i$, should be priced depending (in part) on the aggregate performance of the world's other stock exchanges. 
However the particularity of Eq.(5) is the non-linearity that enters via the theta-functions of $R^{cum}_{ij}$ which ensure an impact across markets only for large, possibly aggregate, price movements.  $R^{cum}_{ij}$ enters the pricing of exchange $i$ with a memory of the past (possible aggregate), price movements of exchange $j$, see first term in Eq.(6).
%``Performance'' here refers to the contribution that exchange $i$ gets from past cumulative returns of a market $j$ once %$R_{ij}^{cum}$  exceeds the threshold $R_C$. 
Similar to the CAPM the sensitivity (called $\beta$ in the CAPM) of asset $i$ to the ``market of exchanges'' is given by a sum of contributions in Eq.(\ref{R_return}) with sensitivity coefficients given by $\alpha_{ij} \beta_{ij}$ now describing the relative impact between stock exchange $i$ and $j$. $\alpha_{ij}$ introduces a weighting in terms of the relative capitalization of the two exchanges $i,j$ (with $\gamma$ setting the scale) and $\beta_{ij}$ introduces a weighting in terms of the difference in time zone (with $\tau$ setting the scale).

The price-quake model can now be written in terms of a cellular automaton as seen for the OFC model. Inserting the impact of the update of stock exchange $i$ at time $t$, Eq.(5), into Eq.(6) one gets:
\ba
\label{R_cum3}    
R_{ij}^{cum} (t) & = &  [1 - \theta (|R_{ij}^{cum} (t-1)| > R_C) ] \times R_{ij}^{cum} (t-1) + \\ 
 & &   {1 \over N_j^*} \sum_{k\neq j}^N \alpha_{j,k} 
 \theta (|R_{jk}^{cum} (t-1)| > R_C) \times R_{jk}^{cum} (t-1) \beta_{j,k} + \eta_j (t)  \ \ \forall j \neq i \nonumber
\ea                                                                                                  
The first term on the right hand side describes the aggregate nature of the stresses imposed on $R_{ij}^{cum}$ which keeps adding up until the value $R_C$ is attained, after which $R_{ij}^{cum}$ is re-initialized (the impact has been priced in, see Eq.'s(5)-(7) ). The second term accounts for the impact across markets whereas the third term takes into account the stresses the price movement of the market $j$ has on all other markets $i$ ($i \neq j$).

It should be noted that another spring-block related approach to financial markets has recently been proposed in S\'andor and N\'eda (2015). In the following we give  a cellular automaton description of $R_{ij}^{cum}$ similar to what was  done for $F_{ij}$ of the OFC model  (see Figure~4). Treat  the open/close of the different markets in sequential order. Assume that the next event to happen at time $t_{present}$ is the close/opening of market $j$, then: \\

{\em
I: If there are no critical stress terms, i.e. $|R_{jk}^{cum}(t_{present} - 1)| < R_c, k \neq j$, then the only impact on $R^{cum}_{ij}(t_{present})$ comes from the local economics news of the exchange $j$, $\eta_j (t_{present})$. If  the stress imposed by $\eta_j$ on all other exchanges $i \neq j$ does not make $R_{ij}^{cum} (t)$ exceed its critical threshold, i.e. $|R_{ij}^{cum} (t-1) + \eta_j(t)| < R_c$  for all $i$, then the system of the stress fields of the exchanges all remain sub-critical and there is presently no impact on the other exchanges.  If on the contrary $|R_{ij}^{cum} (t-1) + \eta_j(t)| > R_c$ the local news on exchange $j$ imposes stresses on the all other exchanges (as the case illustrated in figure~4a) which in turn will have an impact when those exchanges close/open for $t > t_{present}$. \\

II:  If there are critical stress terms, i.e. $|R_{jk}^{cum}(t_{present} - 1)| > R_c, k \neq j$ then the impact on $R^{cum}_{ij}$ comes partly from the local economics news and partly from past price movements of the other exchanges as expressed via $R_{jk}^{cum}(t_{present} - 1)$. This case is illustrated in figure~4b where exchange 2 received a contribution from exchange 1 due to its past price behavior. The impact may then, or may not, propagate further. For example Figure~4b illustrates a case where exchange 2 does not itself incite further perturbations given the initial influence from the exchange 1.
} \\ 
 
It is important to note that the oscillating field in the OFC model is the {\em magnitude} of the force, $F_{i,j}$, on a given block $(ij)$ so the oscillating field in this case is a {\em scalar}. However in the price-quake model the oscillating stress field, $R^{cum}_{ij}$, acts between two stock exchanges $i$ and $j$ therefore in this case the oscillating field is not a scalar but a second order {\em tensor}, just like a stress tensor in Physics. Figure~4 illustrates the dynamics and couplings that exists between the different elements of the stress tensor $R^{cum}_{ij}$. From Figure~4 it is clear that each of the elements $R^{cum}_{ij}$ corresponds to an IAF oscillator since it accumulates stresses up to a point $R_C$ after which the events have been priced in, and the stress tensor element is re-initialized (see e.g. the element $(2,1)$ in Figure~4b). The coupling to the other IAF oscillators is given by the dynamics described from Eq.(8).

%II) The other possibility is that indeed the update of stock exchange $i$ at time $t$ {\em does} lead to $|R_{ij}^{cum}(t)%| > R_c$ from Eq.(9). This scenario is represented in figure~\ref{Fig4}c where the column representing $R_{3j}^{cum}(t)$ is shown in black to show that the column is critical and will in turn impact all the other stock exchanges.

In summary we have seen that in the price-quake model the stress field $R_{ji}^{cum}$ is a second order tensor which acts as an unit IAF oscillator coupled in network of similar IAF oscillators. The units oscillators integrate via Eq.(8), fire when condition II) above is fulfilled and resets to 0 (again via Eq.(6)). It should be noted that an important difference compared to the OFC model is that increments of the stress field $R_{ji}^{cum}$ can take both signs whereas the increments of the force field $F_{ij}$ only took positive values.

\begin{figure}[!htbp]
 \includegraphics[width=14cm]{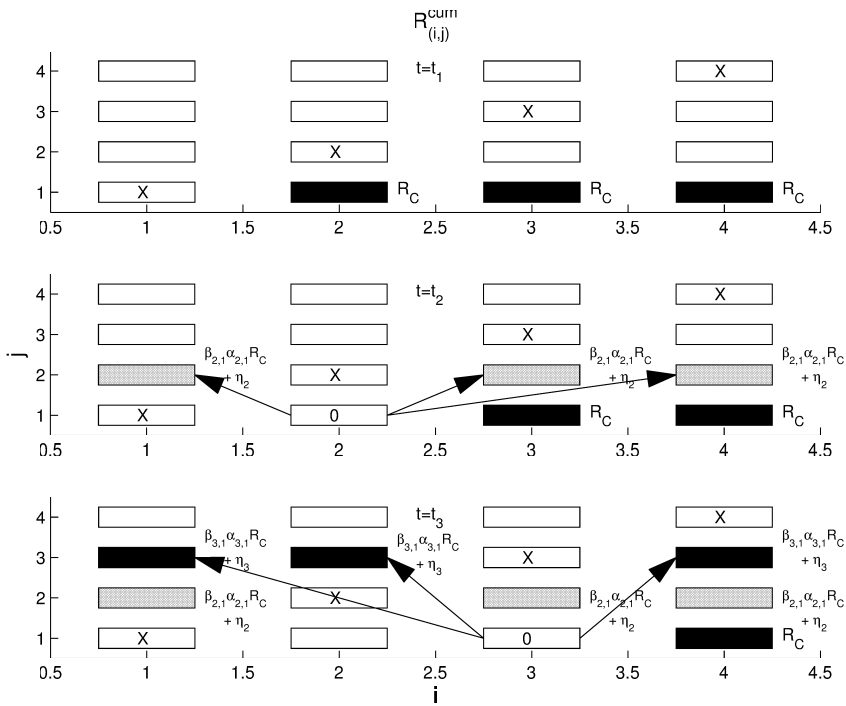}
\caption{
\protect\label{Fig4}
Illustration of the dynamics of $R_{ij}^{cum}$ in the ``price-quake'' oscillator model. a) at $t=t_1$ a large price movement of the index 1 imposes stresses on all the other indices $i \neq 1$ via Eq. (6) with $R_{1i}^{cum} \geq R_c$. 
b) at $t=t_2$ index 2 opens/closes, the impact from index 1 felt by index 2 is priced in
 ($R^{cum}_{21} = 0$)
 and the index 2 in turn impacts the other indices, in this case however without any critical impact on  $R^{cum}_{i2}$. 
c) at $t=t_3$ index 3 opens/closes and the impact from index 1 felt by index 3 is priced in ($R^{cum}_{31} = 0$). 
The index 3 in turn impacts the other indicies in this case a critical impact,   $R_{3i}^{cum} \geq R_c$, which eventually can lead to other future critical impacts.
}
\end{figure}

\section{\large 3. Calibration and empirical findings of the IAF oscillator model}
In Table I is shown the 24 indicies we will use in this study of the IAF oscillator model. The same set of daily open and close were used in (Vitting Andersen et al., (2011)).
 
% TABLE 1 HERE
\begin{table}
\begin{tabular}{lll}
\hline
{Country}&{Stock index: Bloomberg name}\\
\hline
{AUSTRALIA}&{AS30 INDEX}\\
{JAPAN}&{NKY INDEX}\\
{SOUTH KOREA}&{KOSPI INDEX }\\
{CHINA}&{SSE50 INDEX }\\
{HONG KONG}&{ HSI INDEX }\\
{TAIWAN}&{TWSE INDEX }\\
{SINGAPORE}&{FSSTI INDEX }\\
{MALAYSIA}&{FBMKLCI INDEX }\\
{INDONESIA}&{JCI INDEX }\\
{INDIA}&{SENSEX INDEX }\\
{ISRAEL}&{TA-100 INEDX }\\
{EGYPT}&{EGX70 INDEX }\\
{U.K.}&{UKX INDEX }\\
{FRANCE}&{CAC INDEX }\\
{GERMANY}&{DAX INDEX }\\
{SWITZERLAND}&{SMI INDEX }\\
{ITALY}&{FTSEMIB INDEX }\\
{NETHERLANDS}&{AEX INDEX }\\
{AUSTRIA}&{ATX INDEX }\\
{ARGENTINE}&{MERVAL INDEX }\\
{BRASIL}&{IBOV INDEX }\\
{U.S.}&{SPX INDEX }\\
{CANADA}&{SPTSX INDEX }\\
{MEXICO}&{MEXBOL INDEX }\\
\hline
\end{tabular}
\caption{List of Indices }
\label{index}
\end{table}

By constructing the conditional probability that the price movement of a given stock exchange would have the same sign as the open-close price movement of the US stock market, a clear non-linear behavior was identified with random price movements of a given stock exchange (i.e. a conditional probability close to 0.5) following small US open-closes, whereas larger US absolute price movements (say of order 0.02-0.04\%) would lead to a conditional probability close to one for a given market (Vitting Andersen et al., (2011)). Otherwise said, a large movement of the prices in the US markets would most likely be followed by the same tendency for the other exchanges, whereas small price movements in the US market would go unnoticed. The tendency was particular clear for the Asian markets which are all closed during the US open-close whereas the European markets would have had some time to react to initial price movements of the US markets (the European markets close shortly after the US markets open). The same tendency was also found when the conditioning was made on a given price movement of a world index (made of aggregate price movements of the stock exchanges). In addition we have tested the impact of size of capitalization of a given market across markets. This was done by conditioning the size of the price movement for a given markets and see its impact in subsequent price movements of the different stock exchanges. As expected the less important (in terms of capitalization) a market would be, the less pronounced would be the impact across markets even for large price movements.

To summarize: empirical data of 24 of the major stock markets show a clear tendency for small price movements to go unnoticed across markets but an imitation of price movements happens following large price movements, notably for the  markets with the largest capitalization (e.g. the US). Similar effects was seen following aggregate price movements of the ensemble of stock exchanges. Such non-linearity seen in the data has been taken into account in the pricing of the price-quake model via the Theta-function given in Eq.'s(5)-(7).

It should be noted that there are five parameters in the model: $N, Rc, \tau, \gamma$ and the standard deviation of the noise term for exchange $i$, $\sigma_i$. In the following we will take the same sigma for all the exchanges, i.e. we let $\sigma \equiv \sigma_i$, but we have also conducted tests using different $\sigma_i$s to see the impact of heterogeneity of volatility in response to economic news for different stock exchanges. We will comment on such results in the following mostly as additional remarks to the more general findings where $\sigma_i$ is taken the same for all indexes.

In (Vitting Andersen et al. (2011)) maximum likelihood was used to find the optimal parameters describing the pricing via the price-quake model. The values of the maximum likelihood tests gave the values: $\gamma =0.8, \tau=20.0, R_C = 0.03$ and $\sigma^2 = 0.0006$. From Eq.(5) it is seen that the difference $\eta_i (t) = 
R_i (t) - {1 \over N_i^*} \sum_{j\neq i}^N \alpha_{i,j} 
\theta (|R_{ij}^{cum} (t-1)| > R_C) \times R_{ij}^{cum} (t-1) \beta_{i,j}$ should be distributed according to a Gaussian distribution. In figure~5 is shown the difference mentioned, calculated similarly to (Vitting Andersen et al. (2011)), but using another data provider (Bloomberg instead of Yahoo). It can be seen that for the parameter values obtained in the maximum likelihood procedure, the distribution describing the economic news terms, $\eta_i$, is well described by a Gaussian distribution. 

\begin{figure}[!htbp]
%xxx 
 \includegraphics[width=8cm]{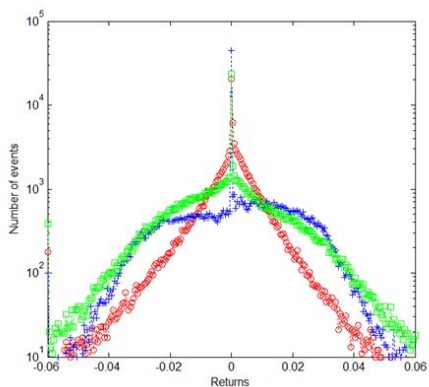}
%FIGURE 5
\caption{\protect\label{Fig5} Number of events (in logarithmic scale) versus return for three different quantities: red circles observed returns $R_{i}$, green squares ${1 \over N_i^*} \sum_{j\neq i}^N \alpha_{i,j} 
\theta (|R_{ij}^{cum} (t-1)| > R_C) \times R_{ij}^{cum} (t-1) \beta_{i,j}$ and blue pluses the difference $\eta_{i}$.}
\end{figure}

\subsection{\large 3.1 Definitions and measurements of price-quakes}
Unlike most other pricing models in finance which are linear, the non-linear properties of our model has the advantage of enabling a precise way of identifying cause and effect in price movements across markets. Specifically we will in the following identify when a price movement in a given market is the cause of later price movements in other markets. Such an identification  will allow us to define a "price-quake" which is a series of price movements caused by the initial price movements in either one or several stock indices. 

In many of the existing IAF oscillator models one study the propagation of avalanches following how the critical amplitude of the oscillators spread through out the network of oscillators see e.g. figure~3 for the OFC model. As mentioned beforehand our unit IAF oscillator is the tensor field $R^{cum}_{ij}$. However from a practical point of view the dynamics of $R^{cum}_{ij}$ is in itself less interesting compared to the price dynamics of the index. We will therefore in the following concentrate on the impact of the price dynamics directly between the different stock indices.To do so we first need to introduce some few definitions. \\

{\em
{\bf Definition 1:} (critical stock index) \\

A stock index $i$ is called critical positive (respectively critical negative) at time $t$ with respect to stock index $j$ if and only if:
\begin{itemize}
\item the stock index $i$ opens/closes at time $t$
\item $R^{cum}_{ij} > R_C$ (respectively $R^{cum}_{ij} < R_C$)
\end{itemize}
}

In order to define a price-quake we need a concept which clearly defines when price movements have an impact from one stock index to another, and vice verse, which price movements in a stock index is influenced via price movements in  another stock index. Given the structure of the pricing formula Eq.(5), as will be seen, it will allow for a clear ``finger-print'' concerning cause and effect of mutual price movements of stock indices.  

We will distinguish between two different cases which we call respectively ``Single Index Price-Quake'' (SIPQ) and ``Cloud Index Price-Quake'' (CIPQ). In the SIPQ case  we follow the perturbation of price movements from one stock index to another. In the CIPQ we instead consider all those stock indices which are influenced by not just one but in principle several stock indices. Before giving the two different definitions we however first need to be clear about what it means to be {\bf influenced} by an index or {\bf impacting} an index.\\ 

{\em
{\bf Definition 2:} (stock index $i$ is {\bf singly influenced} by  stock index $j$ -  stock index $j$ {\bf singly impacting} stock index $i$)

A stock index $i$ has a price dynamics at time $t$ which is {\bf singly influenced} by the price dynamics of stock index $j$, stock index $j$ {\bf singly impacts} stock index $i$ if and only if 
\begin{itemize}
\item The stock index $i$ either opens or closes at time $t$ 

\item $ |R^{cum}_{ij} (t-1) +  \alpha_{i,j} R_{ij}^{cum} (t-1) \beta_{i,j} | > R_C $ where $t-1$ is the previous time for which exchange $i$ opened or closed

\item $ t-1 \leq \tau < t$

\item the stock index $j$ is critical at time $\tau$

\end{itemize}             
}

The reason behind the definition 2 is that we know that only those exchanges $j$ which gives a contribution in the pricing formula Eq.(5) via the term $ \alpha_{i,j} 
\theta (|R_{ij}^{cum} (t-1)| > R_C) \times R_{ij}^{cum} (t-1) \beta_{i,j}$ can influence stock exchange $i$ at time $t$. 
We can now define a single index price-quake as follows: \\

{\em
{\bf Definition 3:} (SIPQ) \\

A single index price-quake begins at time $t_s$, is caused by stock index $i$, and lasts for $T$ time steps if and only if 
\begin{itemize}
\item the stock index $i$ is critical at time $t_s$ and is not influenced by any other stock index at time $t_s$
\item each stock index which at time $\tau > t_s$ is {\bf impacted} by index $i$ becomes part of the set of the single index price quake. Future generations that are {\bf impacted} by the set of indices of the quake themselves become part of the quake. 
\item the quake stops at time $t_s + T$ where $t_s + T$ is the last time an index in the quake {\bf impacts} another index.
\end{itemize}
}
 
In order to consider the case of possible multiple influence/impacts we define: \\ 

{\em
{\bf Definition 4:} (stock index $i$ is {\bf multiply influenced} by a set of stock indices $C$ -  the set of stock indices $C$  {\bf multiply impacting} stock index $i$)

A stock index $i$ has a price dynamics at time $t$ which is {\bf multiply influenced} by the price dynamics of a set of stock indices  $C$,  the stock indicies $C$ {\bf multiply impacts} stock index $i$ if and only if 
\begin{itemize}
\item The stock index $i$ either opens or closes at time $t$ 

\item $ |R^{cum}_{ij} (t-1) +  \sum_{j \in C} \alpha_{i,j} R_{ij}^{cum} (t-1) \beta_{i,j} | > R_C $ where $t-1$ is the previous time for which exchange $i$ opened or closed

\item $ t-1 \leq \tau < t$

\item the stock index $j$ is critical at time $\tau$

\end{itemize}             
}

The definition of a cloud index price-quake then follows along the same lines as that of the single index price-quake: \\

{\em
{\bf Definition 5:} (CIPQ) \\

A cloud index price-quake begins at time $t_s$, is caused by stock index $i$, and lasts for $T$ time steps if and only if 
\begin{itemize}
\item the stock index $i$ is critical at time $t_s$ and is not influenced by any other stock index at time $t_s$
\item each stock index which at time $\tau > t_s$ is {\bf multiply impacted} by a set of indices $C$ becomes part of the set of the cloud index price quake. Future generations that are {\bf multiply impacted} by the set of indices of the quake themselves become part of the quake. 
\item the quake stops at time $t_s + T$ where $t_s + T$ is the last time an index in the quake {\bf multiply impacts} another index.
\end{itemize}
}

\begin{figure}[!htbp]
\includegraphics[width=10cm]{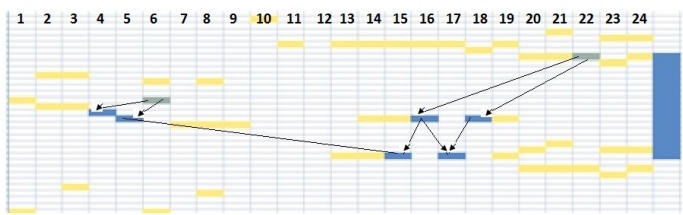} %xxx
\caption{\protect\label{aval2}    
An example of a SIPQ: the 24 indicies are represented as columns whereas the rows represent the hours in a day increasing from top to bottom. Since we only consider opening/closure times there are 24 markets and 48 events per day. Markets that open or close at the same hour are considered a simultaneous event with no interaction between such markets.  In this way for every day the simulation of the model doesn't manage 48 different times but only, in average, 16.  A positive critical return of a stock index is represented by a blue or grey rectangle, whereas stock indicies which are not critical  are represented by a yellow rectangle.  Finally, the blue line in the last column indicates the presence of a positive avalanche. A critical stock index can be critical because another critical stock index impacted it (blue rectangle) or critical caused by local news, i.e. not impacted by another index (gray rectangle).
.}
\end{figure}

\begin{figure}[!htbp]
%xxx
\includegraphics[width=10cm]{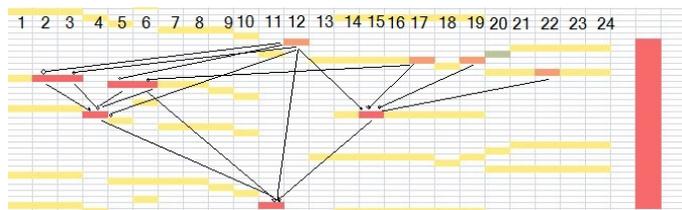}
\caption{\protect\label{aval4} 
An example of a CIPQ: the negative critical stock indicies are represented by a red or orange rectangle, whereas the stock indicies which are not critical are represented by a yellow rectangle.  The last column of the red line indicates the presence of a negative price-quake. As for the single index price-quakes a critical stock index can be negative critical because of an another critical stock index has impacted it (red rectangle) or due to a local cause without influence from the network  (orange rectangle).
The price-quake is composed of 11 nodes in which there are four nodes (orange rectangles) which are not influenced by other nodes: 12,17,19 and 22. Node 20 is critical but does not influence other nodes and is therefore not part of the avalanche. Consequently  it is represented in gray}

\end{figure}

Figure~\ref{aval2} gives an illustration of a SIPQ, whereas and figure ~\ref{aval4} gives an illustration of a CIPQ.

Figure  \ref{aval2}  shows a positive avalanche composed of 8 nodes in which there are two nodes (gray rectangles) 6 and 22 which are not influenced by any nodes but do impact other nodes.  Such nodes are interesting because they play the role of the source for avalanches. In terms of graph-theory these nodes have a zero in-degree, see discussion in the following section. 

As an additional test of our model we then compared simulated SIPQ distributions with the empirical observed. To do so we first  performed computer simulations of the price-quake model using the parameter values as found in (Vitting Andersen et al. (2011)). That is we generated normal distributed random variables $\eta_i$ for each stock exchange (using same distribution for all $i$) and updated the tensor stress field $R_{ji}^{cum}$ via Eq.(9). After having ensured that the system had entered a steady state we then measured the avalanche size distribution and avalanche time distribution shown in figure~8a respectively figure~9a. We then compared these results to the avalanche distributions obtained from the real data of the 24 stock exchanges. These results appear in figure~8b respectively figure~9b. It should be noted that to obtain the distributions for the real data the stress field $R_{ji}^{cum}$ was not obtained from Eq.(9) but instead via Eq.(7). As can be seen by comparing figure~8a and figure~8b, respectively figure~9a and figure~9b, there is a nice agreement between the simulated results which had as only input the random normal distribution of local returns and the real data using the returns of the stock exchanges.

\begin{figure}[!htbp]
%xxx
 \includegraphics[width=7cm]{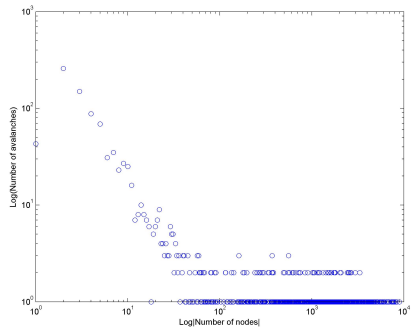}
%xxx 
\includegraphics[width=7cm]{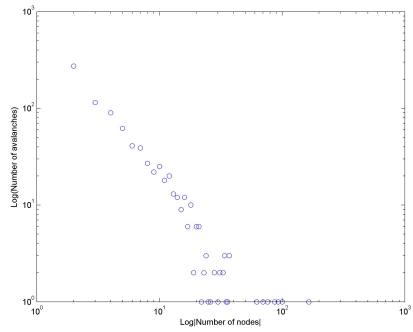}
%FIGURE 8a FIGURE 8b
\caption{\protect\label{Dist_nodes}    
Probability distributions of the size of SIPQ avalanches: a) simulated data, b) empirical data
.}
\end{figure}

\begin{figure}[!htbp]
%xxx 
\includegraphics[width=7cm]{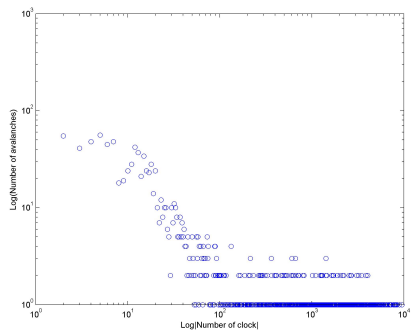}
%xxx 
\includegraphics[width=7cm]{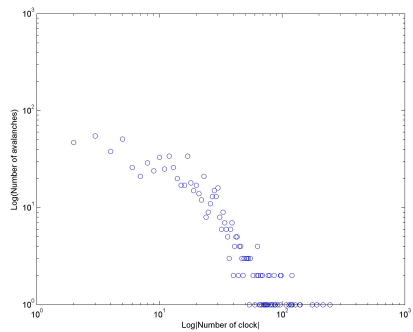}
%FIGURE 9a FIGURE 9b
\caption{\protect\label{Dist_times}
Probability distributions of the time duration of SIPQ avalanches: a) simulated data, b) empirical data
.}
\end{figure}

\subsection{\large 3.2 Empirical findings}

Having validated the parameter values of the IAF model we then use those values to obtain the SIPQ and CIPQ statistics in a time interval over the period from 1/1/2000 to 31/12/2008. Table~II and III give the main statistics of the of both types of avalanches present in the network of stock exchanges of the given time period. As expected from the definitions of the different types of avalanches one observes in average longer lasting avalanches involving more indicies for type CIPQ compared to SIPQ. However as can be seen from figure~8-9 the variation in the number of nodes involved in an avalanche is large as illustrated by the power law shape of the pdfs.  

%TABLE 2 HERE
\begin{table}[!htbp]
\begin{tabular}{cccc}
\hline
{}&{Number of SIPQ}&{    Averaged number of indicies involved}&{    Averaged duration (days)}\\
\hline
{Negative SIPQ}&{337}&{8.32}&{1.39}\\
{Positive  SIPQ}&{515}&{5.27}&{1.22}\\
{Total SIPQ}&{852}&{6.46}&{1.28}\\
\hline
\end{tabular}
\caption{ SIPQ Statistics}
\label{SIPQ_st}
\end{table}

%TABLE 3 HERE
\begin{table}[!htbp]
\begin{tabular}{cccc}
\hline
{}&{Number of CIPQ}&{    Averaged number of indicies involved}&{    Averaged duration (days)}\\
\hline
{Negative CIPQ}&{334}&{8.72}&{1.45}\\
{Positive  CIPQ}&{497}&{6.13}&{1.33}\\
{Total CIPQ}&{831}&{7.17}&{1.38}\\
 \hline
\end{tabular}
\caption{CIPQ Statistics}
\label{CIPQ_st}
\end{table}

Since a SIPQ describes the case where one market impacts directly another (see definition 3)  we will in the following limit our discussion to this case due to its interpretation which is more straight forward compared to a CIPQ where several markets can be involved in the impact of a given market. It should be noted however that the statistics of the two types of avalanches are quite similar since indeed a SIPQ is a special case of a CIPQ. 
The Table 4
%\ref{S_activity} 
shows the number of times in which a stock index is critical. One can distinguish between a critical stock index which is influenced (stock index  4,5,15,16,17, and 18 in Figure \ref{aval2}), a critical stock index that impacts another index but is not influenced itself (stock index 6 and 22 in Figure \ref{aval2}) and stock index which is critical but does not belonging to a price-quake (stock index 20 Figure \ref{aval4}). It can be seen that the markets which are the most volatile, e.g. South Korea, Taiwan and Argentine, are also the markets which are most often critical. But being critical does not mean impacting other markets since it depends several factors such as for example the capitalization of the market.  It is not surprising to see that the market with the largest capitalization, the US, is also the market which is the least times influenced in a SIPQ. 

% TABLE 4 HERE
\begin{table}
\begin{tabular}{lcr}
{~~~~~~~~~~~~~~~~~~~~~~~~~~~~~~~~~~critical positive}&{~~~~~~~~~~~~~~~~~~~~~~critical negative~~~~~~~~}&{~~~\scriptsize{Total amount}}
\end{tabular}\\
\begin{tabular}{llllllllr}
\hline
{}&{\scriptsize{{\bf not} influenced }}&{\scriptsize{{\bf not} influenced }}&{\scriptsize{  influenced }}&{\scriptsize{{\bf not} influenced }}&{\scriptsize{{\bf not} influenced }}&{\scriptsize{  influenced }}&{}\\
{}&{\scriptsize{{\bf not}  impacting}}&{\scriptsize{impacting}}&{\scriptsize{ }}&{\scriptsize{{\bf not}  impacting}}&{\scriptsize{ impacting}}&{\scriptsize{ }}&{}\\
\hline
{\footnotesize{AUSTRALIA}}&{17}&{19}&{44}&{6}&{8}&{52}&{146}\\
{\footnotesize{JAPAN}}&{47}&{37}&{72}&{50}&{45}&{70}&{321}\\
{\footnotesize{SOUTH KOREA}}&{74}&{59}&{134}&{66}&{57}&{128}&{518}\\
{\footnotesize{CHINA}}&{84}&{49}&{51}&{80}&{30}&{60}&{354}\\
{\footnotesize{HONG KONG}}&{34}&{41}&{114}&{28}&{40}&{116}&{373}\\
{\footnotesize{TAIWAN}}&{66}&{64}&{100}&{74}&{58}&{103}&{465}\\
{\footnotesize{SINGAPORE}}&{9}&{13}&{109}&{9}&{5}&{122}&{267}\\
{\footnotesize{MALAYSIA}}&{22}&{22}&{47}&{14}&{12}&{58}&{175}\\
{\footnotesize{INDONESIA}}&{40}&{29}&{118}&{36}&{13}&{100}&{336}\\
{\footnotesize{INDIA}}&{67}&{53}&{141}&{58}&{36}&{118}&{473}\\
{\footnotesize{ISRAEL}}&{24}&{22}&{100}&{24}&{17}&{94}&{281}\\
{\footnotesize{EGYPT}}&{25}&{0}&{81}&{17}&{0}&{56}&{179}\\
{\footnotesize{U.K.}}&{14}&{29}&{59}&{8}&{28}&{70}&{208}\\
{\footnotesize{FRANCE}}&{28}&{62}&{83}&{30}&{61}&{90}&{354}\\
{\footnotesize{GERMANY}}&{31}&{66}&{69}&{30}&{69}&{65}&{330}\\
{\footnotesize{SWITZERLAND}}&{20}&{35}&{88}&{20}&{39}&{82}&{284}\\
{\footnotesize{ITALY}}&{12}&{26}&{82}&{10}&{26}&{100}&{256}\\
{\footnotesize{NETHERLANDS}}&{11}&{22}&{129}&{12}&{18}&{148}&{340}\\
{\footnotesize{AUSTRIA}}&{46}&{25}&{65}&{22}&{24}&{66}&{248}\\
{\footnotesize{ARGENTINE}}&{116}&{54}&{92}&{97}&{37}&{94}&{490}\\
{\footnotesize{BRASIL}}&{80}&{69}&{88}&{75}&{54}&{77}&{443}\\
{\footnotesize{U.S.}}&{40}&{38}&{23}&{48}&{45}&{15}&{209}\\
{\footnotesize{CANADA}}&{29}&{24}&{65}&{22}&{28}&{57}&{225}\\
{\footnotesize{MEXICO}}&{45}&{30}&{102}&{23}&{17}&{95}&{312}\\
\hline
\end{tabular}
\caption{Activity in SIPQ: number of times in which a node is critical}
\label{S_activity}	
\end{table}

In order to get a more qualitative understanding one can analyze the different roles that the different markets play. We show in Table 5
%\ref{S_influence} 
the percentage of time a critical market is influenced/impacted. It should be noted that in this table one can read off from the second column where the price-quakes start the most frequently conditioned that a market is critical. Again we can see the important role of the US market as a source for avalanches, but also larger markets like Germany and France are seen to often impact other markets once they attain criticality themselves. 
The role as the source for avalanches is particularly important: when in a market initiates a large number of price-quake  it  means that it is 
able to influence the network.  Vice versa, the market in which no price-quake starts plays a secondary role in the price dynamics of the network.

%TABLE 5 HERE 
 
\begin{table}
\begin{tabular}{|l|c|c|c|}
\hline
{}&{\scriptsize{ critical not influenced}}&{ \scriptsize{critical not influenced }}&{\scriptsize{critical} }\\
{}&{\scriptsize{ and not impacting}}&{ \scriptsize{but  impacting }}&{\scriptsize{ influenced} }\\
\hline
{AUSTRALIA}&{15.1\%}&{16.5\%}&{68.9\%}\\
{JAPAN}&{27.6\%}&{24.1\%}&{50.1\%}\\
{SOUTH KOREA}&{26.8\%}&{25.3\%}&{60\%}\\
{CHINA}&{43.1\%}&{19.1\%}&{38.8\%}\\
{HONG KONG}&{14.4\%}&{20.8\%}&{72.7\%}\\
{TAIWAN}&{30.8\%}&{28.8\%}&{47.4\%}\\
{SINGAPORE}&{6.7\%}&{6.7\%}&{92\%}\\
{MALAYSIA}&{18.9\%}&{22.4\%}&{62.1\%}\\
{INDONESIA}&{22.5\%}&{13.7\%}&{68.7\%}\\
{INDIA}&{26.5\%}&{20.5\%}&{63.1\%}\\
{ISRAEL}&{15.6\%}&{16.7\%}&{72.1\%}\\
{EGYPT}&{20.6\%}&{2.2\%}&{77\%}\\
{U.K.}&{7.2\%}&{17.7\%}&{75\%}\\
{FRANCE}&{16.7\%}&{33.2\%}&{57.9\%}\\
{GERMANY}&{18\%}&{38.6\%}&{46.1\%}\\
{SWITZERLAND}&{13.7\%}&{26.7\%}&{65\%}\\
{ITALY}&{6.1\%}&{19.4\%}&{80.1\%}\\
{NETHERLANDS}&{5\%}&{11.7\%}&{91.4\%}\\
{AUSTRIA}&{24.1\%}&{20.5\%}&{55.2\%}\\
{ARGENTINE}&{40.3\%}&{19.8\%}&{43.3\%}\\
{BRASIL}&{31.4\%}&{27.6\%}&{41.1\%}\\
{U.S.}&{26.4\%}&{35.5\%}&{38.4\%}\\
{CANADA}&{18.5\%}&{22.8\%}&{65.7\%}\\
{MEXICO}&{18\%}&{17.6\%}&{64.6\%}\\
\hline
\end{tabular}
\caption{Different roles of the nodes in a SIPQ: impacting or influenced?}
\label{S_influence}	
\end{table}

Adding further network analysis one can compute the average in-degree and out-degree over all the avalanches.
For each node one obtains the results shown in Table 6. %~\ref{SNA_degree}.

%TABLE 6 HERE
 \begin{table}
 \begin{tabular}{ccc}
 \hline
  {\scriptsize{~~~~~~~~~~~~~~~~~~~~~~~~~~~positive SIPQ}}&{\scriptsize{~~~~~~~~~~~~~negative SIPQ}}&{\scriptsize{~~~~~~~positive and negative SIPQ}}\\
 \end{tabular}\\
 \begin{tabular}{lccccccccc}
 \hline
 {}&\rotatebox{90}{\tiny{in-degree}}&\rotatebox{90}{\tiny{out-degree}}&\rotatebox{90}{\tiny{$\Delta$(IN-OUT)}}&\rotatebox{90}{\tiny{in-degree}}&\rotatebox{90}{\tiny{out-degree}}&\rotatebox{90}{\tiny{$\Delta$(IN-OUT)}}&\rotatebox{90}{\tiny{in-degree}}&\rotatebox{90}{\tiny{out-degree}}&\rotatebox{90}{\tiny{$\Delta$(IN-OUT)}}\\
 \hline
 {\tiny{AUSTRALIA  }}&{0.31}&{0.26}&{0.05}&{0.62}&{0.62}&{0}&{0.47}&{0.44}&{0.03}\\
 {\tiny{JAPAN  }}&{0.37}&{0.56}&{-0.19}&{0.6}&{1.28}&{-0.68}&{0.49}&{0.92}&{-0.43}\\
 {\tiny{SOUTH KOREA  }}&{0.64}&{0.8}&{-0.15}&{1.14}&{1.4}&{-0.26}&{0.89}&{1.1}&{-0.21}\\
 {\tiny{CHINA  }}&{0.21}&{0.4}&{-0.18}&{0.52}&{0.62}&{-0.1}&{0.37}&{0.51}&{-0.14}\\
 {\tiny{HONG KONG  }}&{0.57}&{0.76}&{-0.19}&{0.99}&{1.59}&{-0.6}&{0.78}&{1.18}&{-0.39}\\
 {\tiny{TAIWAN  }}&{0.6}&{0.57}&{0.04}&{1.2}&{1.1}&{0.11}&{0.9}&{0.83}&{0.07}\\
 {\tiny{SINGAPORE  }}&{0.73}&{0.33}&{0.41}&{1.78}&{0.61}&{1.17}&{1.25}&{0.47}&{0.79}\\
 {\tiny{MALAYSIA}}&{0.2}&{0.13}&{0.07}&{0.72}&{0.3}&{0.42}&{0.46}&{0.22}&{0.25}\\
 {\tiny{INDONESIA}}&{0.68}&{0.32}&{0.35}&{1.53}&{0.5}&{1.03}&{1.1}&{0.41}&{0.69}\\
 {\tiny{INDIA}}&{0.65}&{0.65}&{0}&{1.14}&{1.2}&{-0.06}&{0.89}&{0.93}&{-0.03}\\
 {\tiny{ISRAEL}}&{0.74}&{0.25}&{0.49}&{1.51}&{0.44}&{1.07}&{1.12}&{0.34}&{0.78}\\
 {\tiny{EGYPT}}&{0.58}&{0.03}&{0.55}&{1.04}&{0.03}&{1.01}&{0.81}&{0.03}&{0.78}\\
 {\tiny{UK}}&{0.28}&{0.49}&{-0.21}&{0.75}&{1.13}&{-0.38}&{0.52}&{0.81}&{-0.3}\\
 {\tiny{FRANCE}}&{0.45}&{0.74}&{-0.3}&{0.85}&{1.52}&{-0.68}&{0.65}&{1.13}&{-0.49}\\
 {\tiny{GERMANY}}&{0.28}&{0.74}&{-0.46}&{0.52}&{1.46}&{-0.94}&{0.4}&{1.1}&{-0.7}\\
 {\tiny{SWITZERLAND}}&{0.52}&{0.54}&{-0.03}&{0.89}&{1.15}&{-0.26}&{0.7}&{0.84}&{-0.14}\\
 {\tiny{ITALY}}&{0.51}&{0.51}&{0}&{1.13}&{1.07}&{0.07}&{0.82}&{0.79}&{0.03}\\
 {\tiny{NETHERLANDS}}&{0.74}&{0.58}&{0.17}&{1.47}&{1.36}&{0.12}&{1.11}&{0.97}&{0.14}\\
 {\tiny{AUSTRIA}}&{0.31}&{0.21}&{0.1}&{0.67}&{0.48}&{0.19}&{0.49}&{0.35}&{0.15}\\
 {\tiny{ARGENTINE}}&{0.44}&{0.44}&{0}&{0.83}&{0.91}&{-0.07}&{0.64}&{0.67}&{-0.04}\\
 {\tiny{ BRAZIL}}&{0.35}&{0.7}&{-0.34}&{0.64}&{1.1}&{-0.46}&{0.5}&{0.9}&{-0.4}\\
 {\tiny{US}}&{0.12}&{0.48}&{-0.37}&{0.1}&{1.04}&{-0.94}&{0.11}&{0.76}&{-0.65}\\
 {\tiny{CANADA}}&{0.32}&{0.37}&{-0.05}&{0.53}&{0.87}&{-0.34}&{0.42}&{0.62}&{-0.19}\\
 {\tiny{MEXICO}}&{0.53}&{0.28}&{0.24}&{1.03}&{0.45}&{0.57}&{0.78}&{0.37}&{0.41}\\
 \hline
 {\tiny{average}}&{0.46}&{0.46}&{0}&{0.93}&{0.93}&{0}&{0.69}&{0.69}&{0}\\
 \hline
 \end{tabular}
 \caption{Average in-degree and out-degree in a SIPQ}
 \label{SNA_degree}
 \end{table}
 
We observe that:
\begin{itemize}
\item The number of links is higher in the negative avalanches meaning that contagious negative price movements tend to include more markets compared to the spreading of positive price movements.
\item It is possible to compute a balance for each node ($\Delta$(in-degree)$-$(out-degree)): 
if it is positive we have a node that is more impacted by other markets than it influences other markets, if it is negative it is the other way around.  
Again this measure reveals the dominant role of strong markets as 
US and Germany and, on the other hand, the completely negligible role in the network of markets such  as for example Egypt, Israel or Singapore.
\end{itemize}

An interesting property of our model is that it allows us to identify those markets which are the main sources for the contagion described via the avalanches. One way to illustrate this is to consider 
a market $i$ in which a SIPQ starts: the market $i$ becomes critical, not necessarily because the network influences it, 
but because of local news ($\eta_{i}$), or a series of local news or impacts from the network that accumulates over time.
Therefore we can obtain a ranking among all market in term of source of contagion if we compute, for each market,  the times in which a market is source of a SIPQ and then divide it by the total 
number of SIPQ's. We obtain the ranking shown in  Table 7. %~\ref{S_father}.

%TABLE 7 HERE
\begin{table}
\begin{tabular}{lc}
\hline
{Percentage of times in which in a market starts a SIPQ}&{}\\
\hline
{GERMANIA}&{15.8\%}\\
\hline
{FRANCE }&{14.4\%}\\
\hline
{BRAZIL }&{14.4\%}\\
\hline
{TAIWAN  }&{14.3\%}\\
\hline
{SOUTH KOREA}&{13.6\%}\\
\hline
{ARGENTINE}&{10.7\%}\\
\hline
{INDIA  }&{10.4\%}\\
\hline
{US  }&{9.7\%}\\
\hline
{JAPAN }&{9.6\%}\\
\hline
{HONG KONG }&{9.5\%}\\
\hline
{CHINA }&{9.3\%}\\
\hline
{SWITZERLAND}&{8.7\%}\\
\hline
{UK  }&{6.7\%}\\
\hline
{ITALY }&{6.1\%}\\
\hline
{CANADA }&{6.1\%}\\
\hline
{AUSTRIA }&{5.8\%}\\
\hline
{MEXICO}&{5.5\%}\\
\hline
{INDONESIA}&{4.9\%}\\
\hline
{NETHERLANDS }&{4.7\%}\\
\hline
{ISRAEL }&{4.6\%}\\
\hline
{MALAYSIA}&{4.0\%}\\
\hline
{AUSTRALIA}&{3.2\%}\\
\hline
{SINGAPORE}&{2.1\%}\\
\hline
{EGYPT}&{0\%}\\
\hline
\end{tabular}
\caption{The ranking of the markets in which a SIPQ starts}
\label{S_father}
\end{table}

A bit to our surprise the US in this case doesn't appear to be the main source of contagion, rather the European markets like Germany and France take the leading role when described by this measure. One way to proceed further along such kind of analysis is to 
consider not just the tendency of a given market to be the source of contagion but also how far the contagion will spread?
In this regard we have computed the average number of markets belonging to an avalanche 
that begins in a given market $i$. The results are given in Table 8. %~\ref{lengSNA}. 
Considering Table 8 %~\ref{lengSNA} 
again emphasizes the role of the US market notably with respect to positive contagion of positive market returns. Also the U.K. appear to be a very dominant market for positive market returns something which was less clear from the other statistics we have considered so far. Quite striking is also the large impact that the Swiss market have but on the negative influence across the network. It should be noted that the apparent impact of the Dutch market is an artifact since this market open/closes on hour before the other European markets and so is the first market to incorporate news that happened in the Asian/ North American markets. This creates a ``spurious'' effect where the Dutch market appear responsible for impacts on the European markets for price movements that happened elsewhere.

%TABLE 8 HERE 

\begin{table}
\begin{tabular}{c}
{Average number of nodes belonging to an avalanche that starts from a market}\\
\end{tabular}
\begin{tabular}{lrrr}
\hline
{}&{Positive SIPQ}&{Negative SIPQ}&{SIPQ}\\
\hline
{AUSTRALIA  }&{12.3}&{7.5}&{9.9}\\
{JAPAN  }&{9.4}&{5.6}&{7.5}\\
{SOUTH KOREA  }&{10.5}&{9.4}&{10}\\
{CHINA  }&{5.9}&{9.2}&{7.5}\\
{HONG KONG  }&{8.6}&{7.1}&{7.8}\\
{TAIWAN  }&{8}&{8.6}&{8.3}\\
{SINGAPORE  }&{8.7}&{5}&{6.8}\\
{MALAYSIA}&{4.5}&{5.2}&{4.8}\\
{INDONESIA}&{5.8}&{8.2}&{7}\\
{INDIA}&{9.2}&{6.8}&{8}\\
{ISRAEL}&{8.8}&{5.6}&{7.2}\\
{EGYPT}&{0}&{0}&{0}\\
{UK}&{15.2}&{5.1}&{10.2}\\
{FRANCE}&{12.8}&{6.8}&{9.8}\\
{GERMANY}&{10.8}&{7.5}&{9.1}\\
{SWITZERLAND}&{10.2}&{12}&{11.1}\\
{ITALY}&{11.3}&{6.2}&{8.7}\\
{NETHERLANDS}&{10.5}&{19}&{14.8}\\
{AUSTRIA}&{9}&{5.8}&{7.4}\\
{ARGENTINE}&{7.4}&{7.5}&{7.5}\\
{BRAZIL}&{9.8}&{6.8}&{8.3}\\
{US}&{14}&{7.3}&{10.6}\\
{CANADA}&{7.3}&{7.3}&{7.3}\\
{MEXICO}&{9.3}&{9.3}&{9.3}\\
\hline
\end{tabular}
\caption{The number of markets, in average, impacted by a SIPQ starting from a given market $i$}
\label{lengSNA}
\end{table}

\clearpage

\section{\large 4. Conclusion}
We have introduced a IAF oscillator model to describe the pricing in the worlds's network of stock exchanges. It has been shown how contagion within such a model can be understood as synchronization of the network of
stock exchanges as a whole, caused in a large part of the exchanges which adjust their “rhythms” (by pricing in
price movements of the other exchanges) thereby producing a global aggregate signal. 
One of the main features
of our model is a separation of time scales with a slow price dynamics due to economic fundamentals for a given
country, and a fast price dynamics due to impact across markets. The characteristic non-linear price behavior of the IAF  oscillators is supported by empirical data and has a behavioral origin. One advantage of the IAF dynamics is that it enables for a direct identification of cause and effect of price movements. Our methodology of identifying cause and effect via avalanche price dynamics combined with network analysis has enable us to identify the most relevant nodes with respect to onset of contagion in the network of stock exchanges.

%\section{\large Acknowledgments}
%The research leading to these results has received funding from the European Union Seventh Framework Programme (FP7-SSH/2007-2013) under grant agreement n° 320270 – SYRTO. The authors thank COST Actions TD1210 and IS1104 for financial support. This work was achieved
%through the Laboratory of Excellence on Financial Regulation (Labex ReFi) supported by PRES
%heSam under the reference ANR-10-LABX-0095. It benefitted from a French government support
%managed by the National Research Agency (ANR) within the project Investissements d’Avenir
%Paris Nouveaux Mondes (invesments for the future Paris-New Worlds) under the reference ANR-
%11-IDEX-0006-02.

%\begin{thebibliography}{}
\section{\large References}
1. Ahmad W., Sehgal S., \&  Bhanumurthy N. R. (2013) Eurozone crisis and BRIICKS stock markets: Contagion or
market interdependence? {\em Economic Modelling 33}, 209225. \\

2. Aloui R., A\"issa M. S. B., $\&$ Nguyen D. K. (2011). Global financial crisis, extreme interdependences, and contagion effects: The role of economic structure? {\em Journal of Banking $\&$ Finance 35} 130–141. \\

3. Bae K.-H., Karolyi G. A., \& Stulz R. M. (2003). A New Approach to Measuring Financial Contagion. {\em The Review of Financial Studies Vol. 16}, No. 3, 717-763. \\

4. Bekaert G., Harvey C. R., \& Ng A. (2003) Market integration and contagion, Working paper 9510 http:/\/\ www.nber.org/\/papers/\/w9510 . \\

5. Caporale C. M., Cipollini A., \& Spagnolo N. (2005). Testing for contagion: a conditional correlation analysis.
{\em Journal of Empirical Finance 12}, 476-489. \\

6. Celik S. (2012). The more contagion effect on emerging markets: The evidence of DCC-GARCH model. {\em Economic
Modelling 29 (5)}, 19461959. \\

7. Chiang T. C., Jeon B. N., \& Li H. (2007). Dynamic correlation analysis of financial contagion: Evidence from
Asian markets, {\em Journal of International Money and finance 26}, 1206-1228. \\

8. Corsetti G., Pericoli M., Sbracia M. (2005). ”Some contagion, some interdependence”: More pitfalls in tests of
financial contagion. {\em Journal of International Money and Finance 24}, 1177-1199. \\

9.  De Bondt W. F. M., Thaler R. (1985). Does the Stock Market Overreact?.{\em The Journal of Finance 40 (3)},
793-805.

10.Dimitriou D., Kenourgios D. , Simos T. (2013). Global financial crisis and emerging stock market contagion:
A multivariate FIAPARCHDCC approach. {\em International Review of Financial Analysis 30}, 4656. \\

11. Fisher I. (1933). The Debt-Deflation Theory of Great Depressions. {\em Econometrica 1}, 337-357. \\

12.Forbes K. J., \& Rigobon R. (2002). No Contagion, Only Interdependence: Measuring Stock Market Comovements. {\em The Journal of Finance, Vol. 57}, No. 5, 2223-2261. \\

13. Gjika D.,  Horvath R. (2013). Stock market comovements in Central Europe: Evidence from the asymmetric
DCC model.{\em Economic Modelling, 33}, 55-64. \\

14. H\"am\"al\"ainen R. P., Luoma J., Saarinen, E. (2013). On the importance of behavioral operational research: The case of understanding and communicating about dynamic systems. {\em European Journal of Operational Research, 228(3)}  623-634. \\

15. Hong Y., Lui Y., \& Wang S. (2009). Granger causality in risk and detection of extreme risk spillover between financial markets.  {\em Journal of Econometrics, 150} , 271-287. \\

16. Jones R. H., Crowell D. H., \& Kapuniai L. E. (1969). Change detection model for serially
correlated data. {\em Psychological Bulletin Vol. 71 (No. 5)}, 352–358. \\

17. Keynes J. M. (1936). {\em The General Theory of Employment, Interest, and Money}. New York: Harcourt, Brace
and Company. \\

18. Lewin T. D., Momen N., Drifdahl S. B., \& Simons D.J. (2000). Change blindness, the
metacognitive error of etimating change detection ability. {\em Vision 7(1,2,3)}, 397–413. \\

19.  Lin S., and Rassenti S. (2012). Are under- and over- reaction the same matter? Experimental evidence. {\em  Journal of Economic Behavior
$\&$ Organization 84},  39-61. \\

20. Lintner J. (1965). The valuation of risk assets and the selection of risky investments in
stock portfolios and capital budgets. {\em Review of Economics and Statistics 47(1)}, 13–37. \\

21. Lynch A. (2000). Thought Contagions in the Stock Market. {\em The Journal of Psychology and Financial Markets 1 (1)}, 10-23. \\

22.  Mensi W., Beljid M., Boubaker A., \& Managi S. (2013). Correlations and volatility spillovers across commodity and stock markets: Linking energies, food, and gold. {Economic Modelling, 32}, 1522. \\

23. Olami Z., Feder H. J. S., \& Christensen K. (1992). Self-organized criticality in a continuous, nonconservative cellular automaton modeling earthquakes. {\em Phys. Rev. Lett. 68}, 1244. \\

24. Rensink R. A. (2002). Change detection. {em Annual Review of Psychology Vol 53}, 245–277. \\

25. Roll, R. (1988). The international cash of October 1987. {\em Financial Analysts Journal 44 (September–October)}, 19–35. \\

26. Rosenblum M., \& Pikovsky A. (2007). Self-Organized Quasiperiodicity in Oscillator Ensembles with Global Nonlinear Coupling. {\em Phys. Rev. Lett.  98}, 064101.\\

27. S\'andor B., \& N\'eda S. (2015). A spring-block analogy for the dynamcis of stock indexes. {\em Physica A Vol. 427} 122-131. \\

28. Sharpe W. F. (1964). Capital asset prices: a theory of market equilibrium under
condition of risk. {\em Journal of Finance 19(3)}, 425–442.\\

29. Shiller, R. J. (1989). {\em Investor behavior in the October 1987 stock market crash: Survey evidence}. In: Shiller, R. J. (Ed.), Market Volatility. MIT Press, Cambridge, pp. 379–402. \\

30. Syllignakis M. N. ,  Kouretas G. P. (2011). Dynamic correlation analysis of financial contagion:
Evidence from the Central and Eastern European markets. {\em International Review of Economics $\&$ Finance, 20 (4)} 717732. \\

31. Treynor J. L. (1999). {\em Toward a theory of market value of risky assets. Asset Pricing
and Portfolio Performance: Models, Strategy and Performance Metrics}. Risk
Books. pp 15–22. \\

32. Vitting Andersen J., \& Nowak A. (2013). {\em An Introduction  to  Socio-Finance}. Berlin: Springer. \\

33. Vitting Andersen J., Nowak A., Rotundo G., Parrot L., \& Martinez S. (2011). ``Price-Quakes'' Shaking the World’s Stock Exchanges. {\em PLoS ONE 6} (11): e26472. Doi:10.1371/journal.pone.0026472. \\
                                                                                                               
34. Wang S.-J., Hilgetag C. C., \& Zhou C. (2011). Sustained Activity in Hierarchical Modular Neural Networks: Self-Organized Criticality and Oscillations. {\em Frontiers in Computational Neuroscience, doi:10.3389/fncom.2011.00030} \\ 

35. Yang J., \& Bessler D. A. (2008). Contagion around the October 1987 stock market crash. {\em European Journal of Operational Research 184} 291–310.
%\end{thebibliography}

\end{document}